\documentclass[conference]{IEEEtran}
\IEEEoverridecommandlockouts
\usepackage{cite}
\usepackage{amsmath,amssymb,amsfonts}
\usepackage{algorithmic}
\usepackage{graphicx}
\usepackage{textcomp}
\usepackage{xcolor}
\usepackage{algorithmic}
\usepackage{algorithm}
\usepackage{booktabs}
\usepackage{multirow}

\def\BibTeX{{\rm B\kern-.05em{\sc i\kern-.025em b}\kern-.08em
    T\kern-.1667em\lower.7ex\hbox{E}\kern-.125emX}}
\begin{document}

\title{PLAA: Packet-level Adversarial Attacks in Network Traffic Detection
%\\ {\footnotesize \textsuperscript{*}Note: Sub-titles are not captured in Xplore and should not be used}
\thanks{This work is supported by National Key Research and Development Program of China (grant No. 2024YFC3015300), Beijing Natural Science Foundation (grant No.4244084), Fundamental Research Funds for the Central Universities (grants No.2024ZCJH04, 530224004, 2024PTB-009), National Natural Science Foundation of China (grants No.62401075, 62394322, 92167204, 62072030).}
}

\author{\IEEEauthorblockN{1\textsuperscript{st} Jinhao You}
\IEEEauthorblockA{\textit{Beijing University of Posts and \ } \\
\textit{Telecommunications}\\
Beijing, China \\
youjinhao@bupt.edu.cn}
\and
\IEEEauthorblockN{2\textsuperscript{nd} Zan Zhou\textsuperscript{*}}
\IEEEauthorblockA{\textit{Beijing University of Posts and \ } \\
\textit{Telecommunications}\\
Beijing, China \\
zan.zhou@bupt.edu.cn
\thanks{*Corresponding author: Zan Zhou (zan.zhou@bupt.edu.cn)}
}
\and
\IEEEauthorblockN{3\textsuperscript{rd} Shujie Yang}
\IEEEauthorblockA{\textit{Beijing University of Posts and \ } \\
\textit{Telecommunications}\\
Beijing, China \\
sjyang@bupt.edu.cn}
\and
\IEEEauthorblockN{4\textsuperscript{th} Yi Sun}
\IEEEauthorblockA{\textit{Beijing University of Posts and \ } \\
\textit{Telecommunications}\\
Beijing, China \\
sybupt@bupt.edu.cn}
\and
\IEEEauthorblockN{5\textsuperscript{th} Lei Zhang}
\IEEEauthorblockA{\textit{Zhongguancun Laboratory} \\
Beijing, China \\
zhanglei@zgclab.edu.cn}
\and
\IEEEauthorblockN{6\textsuperscript{th} Changqiao Xu}
\IEEEauthorblockA{\textit{Beijing University of Posts and \ } \\
\textit{Telecommunications}\\
Beijing, China \\
cqxu@bupt.edu.cn}
}

 \maketitle

\begin{abstract}
Deep neural networks (DNNs) are widely applied in Network-based Intrusion Detection System (NIDS) due to their high accuracy. However, DNNs are highly susceptible to adversarial attacks, which generate malicious traffic to evade NIDS detection. Existing approaches often adapt adversarial attacks from computer vision (CV) tasks to the NIDS domain, overlooking the fundamental differences between CV and NIDS. This results in two major issues: 1) The generated network traffic may become invalid, 2) The generated traffic may lose its original attack semantics. To address these issues, this paper proposes an adversarial attack specifically designed for NIDS. Instead of directly generating flow-level features, our approach incrementally generates packet-level features to construct adversarial traffic. During the generation process, the semantic integrity of the traffic is monitored at each stage, effectively avoiding the issues of invalid traffic and semantic loss observed in existing methods. We evaluate our attack algorithm against current NIDS models using the CIC-UNSW-NB15, CIC-DDoS2019, and CIC-IDS-2017 datasets. The proposed method achieves an average evasion success rate of 92.78\%, while ensuring that the generated adversarial traffic remains semantically consistent with the original malicious traffic.

%深度神经网络（RNN）因其高精度而在基于网络的入侵检测系统（NIDS）中得到了广泛的应用。然而，RNN极易受到对抗性攻击，这些攻击会产生恶意流量以逃避NIDS检测。现有的方法通常将计算机视觉（CV）任务的对抗性攻击适应到NIDS领域，忽略了CV和NIDS之间的根本区别。这导致了两个主要问题：1）生成的网络流量可能会无效，2）生成的流量可能会失去其原始的攻击语义。为了解决这些问题，本文提出了一种专门为NIDS设计的对抗性攻击。它避免了直接生成flow-level的特征，而是通过逐一生成packet-level特征从而生成恶意流量，并在生成的过程中，逐级监控所生成流量的语义，这有效地避免了现有方法中观察到的流量无效和攻击语义丢失的问题。我们在CIC UNSW-NB15，CIC-DDoS2019, CIC-IDS-2017数据集上针对当前的NIDS评估了我们的攻击算法，在规避方面取得了平均92.78%的成功率，并确保所生成的对抗性流量与原有恶意流量相似并保持其攻击语义。

\end{abstract}

\begin{IEEEkeywords}
Network-based Intrusion Detection System, adversarial attack, reinforcement learning, packet-level, attack semantic
\end{IEEEkeywords}

\section{Introduction}
With the development of computer networks, cyberattacks have rapidly evolved, posing serious threats to network systems \cite{aktar2023towards}. Network-based Intrusion Detection System (NIDS) offers an effective solution to such issues. A typical NIDS operates through three main stages: monitoring, detection, and response \cite{corona2013adversarial}. In the monitoring stage, flow-based statistical features of network traffic, such as the average packet length and inter-packet intervals, are collected. During the detection stage, the monitored traffic's flow-based statistical features are analyzed and classified to identify malicious traffic. Finally, based on the classification results, the system takes appropriate response actions to defend against malicious traffic attacks.

Deep Learning (DL), with its exceptionally high accuracy in classification tasks, has been increasingly adopted by a growing number of NIDS in the detection phase \cite{zhou2020endogenous} \cite{lin2023drl}. However, NIDS that utilize DL are highly susceptible to adversarial attacks \cite{szegedy2013intriguing}. Initially, such attacks were primarily applied in the field of computer vision (CV). Attackers would first craft subtle, almost imperceptible perturbations, known as adversarial examples, and add them to images to deceive DNNs. This greatly inspired network attackers, who began attempting to modify or reshape the statistical features of malicious traffic to evade DL-NIDS \cite{zhou2022multi}. For instance, they might add adversarial examples to the statistical features of malicious traffic or generate carefully designed malicious traffic features.

\begin{figure}[h]
	\includegraphics[width=\columnwidth]{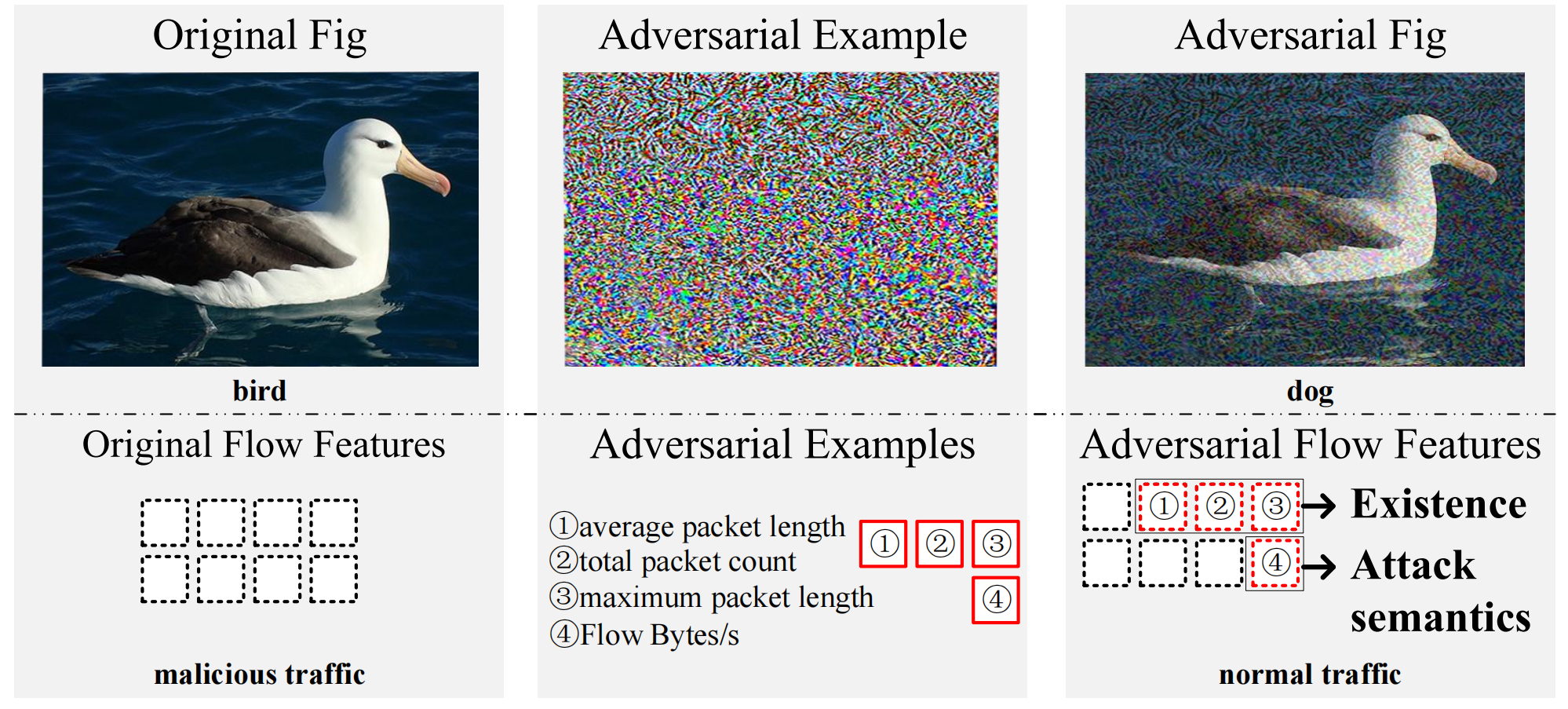}
	\caption{Differences in adversarial attacks between CV and NIDS.} \label{fig1}
\end{figure}

However, most existing adversarial attacks are adaptations from the CV field, often overlooking the fundamental differences between CV and NIDS \cite{he2023adversarial}. In the CV field, attackers can generate arbitrary adversarial examples and apply them to any part of an image. However, adding adversarial examples to the statistical features of network traffic can lead to two key issues, as illustrated in the Fig. \ref{fig1}.

\textbf{Issue of existence}: The statistical features of network traffic are interdependent, and adversarial examples generated by existing attacks often fail to consider these dependencies. Simply adding adversarial examples to the statistical features of network traffic or directly generating malicious traffic using adversarial methods can result in traffic that is unrealistic in practice. For example, in network traffic, features such as average packet length, total packet count, and maximum packet length are interrelated. Modifying the average packet length requires corresponding adjustments to the other statistical features to maintain consistency.

\textbf{Issue of attack semantics}: Taking Denial of Service attack (DoS) as an example, malicious traffic must maintain a certain high value in statistical features such as ``Flow bytes/s" and ``Flow packets/s" in order to achieve its attack semantics. Using traditional adversarial attacks to generate malicious traffic can indeed evade NIDS detection, but its attack semantics are difficult to guarantee.

\begin{figure}[h]
	\includegraphics[width=\columnwidth]{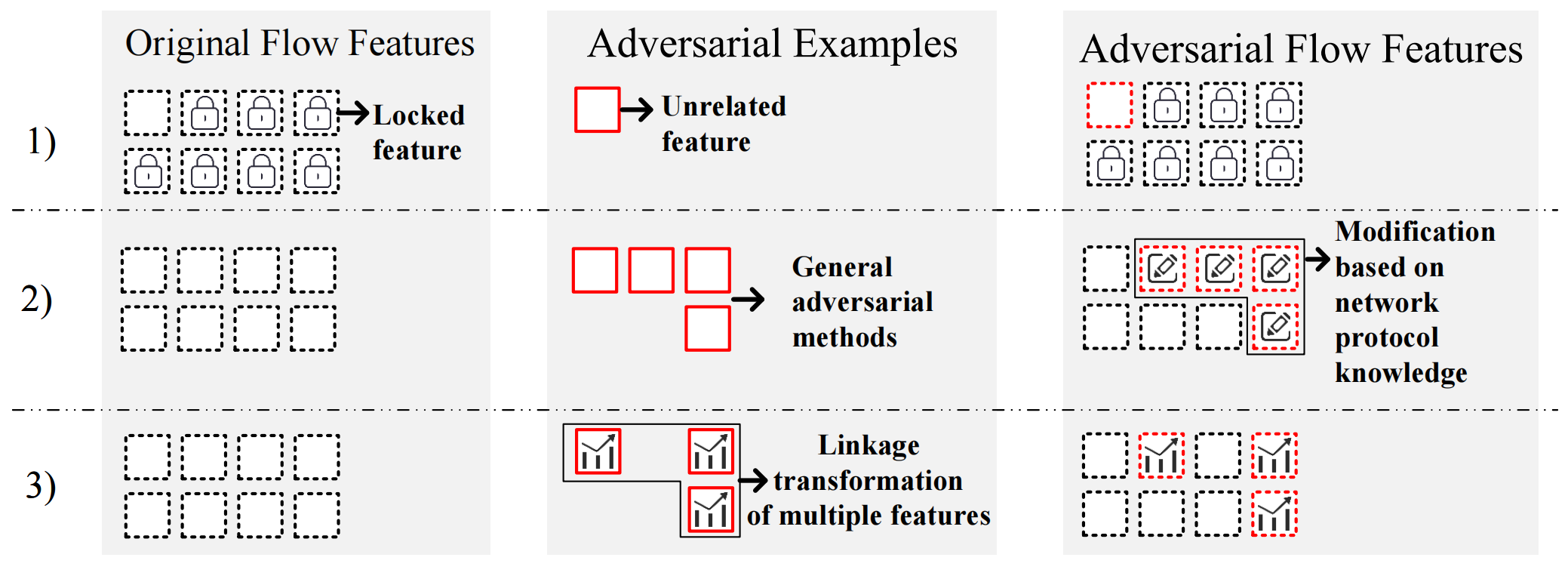}
	\caption{Differences in adversarial attacks between CV and NIDS.} \label{fig2}
\end{figure}

To address these issues, existing studies commonly use three methods as shown in Fig. \ref{fig2}: 1) locking key statistical features of network traffic before generating or modifying malicious traffic, making adjustments to a small number of statistical features that are unrelated to other features to ensure the traffic exists and retains its attack semantics \cite{lin2022idsgan} \cite{sharon2022tantra}; 2) adjusting the malicious traffic based on network protocol knowledge after it has been generated or modified \cite{debicha2023adv} \cite{zolbayar2022generating}; 3) define adversarial modification methods for limited network traffic based on existing network protocol knowledge, in order to achieve linkage transformation of multiple statistical features of malicious traffic \cite{yan2023automatic}. However, these approaches have drawbacks: 1) locking parameters restricts the performance of adversarial attacks; 2) modifying traffic after generation severely impacts its effectiveness; 3) protocol-based adversarial methods require tailored modifications for different network intrusion detection systems (NIDS), demanding significant expertise and repeated testing.

To fundamentally address the issues of existence and attack semantics, this paper adopts a reinforcement learning (RL) framework to generate packet-level features one by one, rather than directly generating flow-level statistical features. Since network traffic consists of a series of packets, directly generating packet-level features and then combining these packets into a flow effectively resolves the existence issue of the generated traffic. Moreover, an attack semantics term is incorporated into the reward function to ensure that the generated network traffic meets the required attack semantics. Notably, under the RL framework, the generated packets are not only focused on the current state of the traffic, such as maximizing the attack semantics or approaching the misclassification boundary of the NIDS. Instead, they prioritize long-term benefits, contributing to the overall effectiveness of the final generated traffic.
Our specific contributions are as follows:

\begin{itemize}
	\item We conduct an in-depth analysis and summary of the limitations of current adversarial network traffic methods in terms of traffic presence and attack semantics.  
	\item  We propose a novel approach that sequentially constructs packet-level feature generation for adversarial traffic, effectively addressing the challenge of traffic presence while preserving the original attack semantics. 
	\item We apply the PLAA method to existing NIDS models and demonstrate its effectiveness in evading NIDS detection while maintaining the original attack semantics across various attack scenarios.
\end{itemize}

In the remainder of this paper, we present the related work in Section \ref{sec2}, define the threat model in Section \ref{sec3}, describe the details of PLAA in Section \ref{sec4}, conduct relevant experiments and analysis in Section \ref{sec5}, and conclude in Section \ref{sec6} with a discussion of future work. 

\addtolength{\topmargin}{0.01in}

\section{Related work}\label{sec2}

% In this section, we will discuss the related work on adversarial attacks against NIDS from two aspects: NIDS and adversarial attacks.

\subsection{NIDS}
Typical NIDS consist of three stages: monitoring, detection, and response. Since this paper focuses solely on the evasion detection of malicious traffic, this section introduces NIDS from the perspectives of 
% monitoring and 
detection.

% \subsubsection{Monitoring}In the monitoring phase, due to the massive volume of network data and the limited processing capacity of monitoring devices, NIDS commonly adopts flow-based feature extraction from network data. Some NIDS have publicly released their feature extractors(such as CICFlowmeter \cite{sharafaldin2018toward}). Specifically, this involves extracting attribute information and statistical information from the same flow. The attribute information provides general details about the flow, such as the source IP and destination IP, and the protocol. In contrast, statistical information measures and aggregates the characteristics of packets within the flow, such as the number of packets, the average length of the packets, and the Inter-Arrival Time  between packets.

% \subsubsection{Detection}
In the detection phase, NIDS utilizes the features obtained during the monitoring phase to identify attacks through traffic classification or anomaly detection. DL used for traffic classification include DNNs, Recurrent Neural Networks (RNNs), and Convolutional Neural Networks (CNNs). DNN, as the most fundamental neural network framework, has been widely applied in NIDS due to its flexible parameters and structure \cite{jia2019network} \cite{otoum2022dl} \cite{zhou2024secfft} \cite{lin2024incentive}. RNN is a type of neural network specifically designed to handle tasks involving sequential features. Subsequently, two improved RNN variants were developed: Long Short-Term Memory (LSTM) and Gated Recurrent Units (GRU). The reason for using this type of neural network to detect network attacks is that network traffic data contains sequential information that is valuable for detection \cite{roopak2019deep}. CNNs focus more on processing data correlations. For instance, in image classification tasks, they extract the relationships between different pixels. In network traffic detection, CNNs are typically used to extract the correlations among different features of the traffic. These extracted correlations are then utilized by RNNs for further analysis \cite{sun2020dl} \cite{zhou2025community}\cite{lin2024scalable}. DL-based anomaly detection NIDS typically use autoencoders (AE) to identify malicious traffic. The encoder and decoder are trained exclusively on benign data. When input traffic features are reconstructed, if the reconstruction error exceeds a predefined threshold, the input is classified as anomalous \cite{mirsky2018kitsune} \cite{nguyen2019gee}.

\subsection{Adversarial attacks}
In adversarial attacks against NIDS, attackers apply an operator $A: \mathbb{R}^{d} \rightarrow \mathbb{R}^{b}$ to add perturbations to network traffic, thereby misleading the NIDS classifier $D: \mathbb{R}^{d} \rightarrow \mathbb{R}$.
\begin{align}\label{eq1}
    X_{adv}=A(X)=X+n \nonumber \\
    D(X_{adv})=C_{t}\neq D(X)=C_{o}
\end{align}
here, $X$ represents the original traffic features, $n$ denotes the perturbations, and the adversarial traffic features $X_{adv}$ will cause the NIDS classifier $D(\cdot )$ to misclassify the malicious traffic, originally belonging to $C_{o}$, as $C_{t}$. Among them, $C_{o}$ represents the true type of traffic (malicious traffic), while $C_{t}$ is the type that the attacker wants $D(\cdot)$ to recognize (benign traffic).

Based on the above definition, it can be seen that existing methods for constructing $x_{adv}$ have two degrees of freedom 
 \cite{wang2024progen}, namely the operator $A(\cdot )$ and the perturbation $n$. We will introduce relevant adversarial attacks based on these two degrees of freedom.

\subsubsection{Adversarial attacks based on perturbations}These attacks follow a relatively similar pattern, assuming that the attacker has knowledge of the parameters and structure of the model used by the NIDS, i.e., they perform white-box (WB) attacks. Adversarial attack algorithms previously applied in the CV field, such as FGSM \cite{ibitoye2019analyzing}, JSMA \cite{huang2019adversarial}, and DeepFool \cite{wang2018deep}, are adapted to the NIDS domain to generate perturbations and add them to the original malicious traffic. However, the malicious traffic generated using these methods often employs similarity constraints, specifically $\ell^{p}$ distance($\| \cdot \|_{p}$), as shown in Equation (\ref{eq2}), where $p$ is the norm.
\begin{align}\label{eq2}
\|\vec{n}\|_{p}=\left(\sum | n_{i} | ^{p}\right)^{\frac{1}{p}}
\end{align}

In the CV field, this effectively measures whether the generated perturbations cause excessive differences between the original image and the adversarial image. Excessive differences may make the image recognizable by humans, leading to the failure of the attack. However, such a similarity measure lacks significance for constraining malicious network traffic. For adversarial malicious traffic, the most critical factor is ensuring that its features comply with network protocol rules rather than its similarity to the original malicious traffic.

\subsubsection{Adversarial attacks based on operators}These attacks primarily focus on enabling the operator to learn the structure of network traffic features and ensuring that the generated traffic's flow features comply with network protocol rules through methods such as traffic feature locking and modifications \cite{lin2022idsgan} \cite{sharon2022tantra} \cite{debicha2023adv} \cite{zolbayar2022generating} \cite{yan2023automatic}. Such methods often utilize feature generation algorithms like Generative Adversarial Networks (GANs). In this process, the generator first generates adversarial traffic based on the original malicious traffic. Then, the discriminator provides feedback to the generator, which undergoes repeated training based on the feedback until the discriminator can no longer distinguish the adversarial traffic. Although such methods can, to some extent, ensure that the generated traffic features meet the requirements of real-world attacks, they still face issues such as low success rates of the generated traffic, loss of original attack semantics, and the need for extensive expert knowledge. 
Although such methods can, to some extent, ensure that the generated traffic features align with the requirements of real-world attacks, they still face challenges such as low success rates of the generated traffic, loss of original attack semantics, and the requirement for extensive expert knowledge.

\subsubsection{Evaluation datasets}Benchmark network traffic datasets are critical for research on NIDS and adversarial attacks. Existing benchmark datasets are designed to simulate network activities generated by organizations consisting of various devices. Commonly used network datasets include NSL-KDD \cite{tavallaee2009detailed}, CIC-UNSW-NB15 \cite{10788064}, and CIC-IDS2017/CSE-CIC-IDS2018 \cite{sharafaldin2018toward}, CIC-DDoS2019 \cite{sharafaldin2019developing}. In this study, we will also validate the proposed method using these datasets.
\addtolength{\topmargin}{0.01in}
\section{Threat model}\label{sec3}

In this study, we define such a threat scenario where an attacker targets a network environment protected by a DL-based NIDS and sends adversarially modified malicious network traffic to evade detection. Based on this scenario, we define the threat model from three aspects: the attacker's objectives, knowledge, and capabilities.

\subsection{Objective}Generate network traffic that can pass NIDS detection, with the traffic needing to possess attack semantics and be implementable by actual network behaviors.

\subsection{Knowledge}The input features of NIDS have always been a focus of interest for attackers. However, in real-world scenarios, the feature extractors used by NIDS during the monitoring phase are often difficult to obtain. Moreover, although many studies claim to replicate NIDS detection models, it is evident that attackers in practical situations would find it challenging to discern the exact structure and parameters of the detection model using such methods. Therefore, assuming that attackers can analyze the detection model to launch a WB attack is unrealistic. In this section, we stipulate that attackers have no knowledge of the features monitored during the NIDS monitoring phase or the structure and parameters of the model used in the detection phase.

\subsection{Capability}Although we cannot know the model's input features or specific structure and parameters, we assume that attackers can access the NIDS and obtain its feedback to carry out a gray-box (GB) attack. This assumption has been implemented in many studies. For example, in \cite{yan2023automatic}, a method was proposed to replicate detection models with minimal access. This method does not aim for structural or parametric similarity between the replicated model and the original model but focuses solely on ensuring that their output behaviors are identical.

\section{Methodology}\label{sec4}

% In this chapter, we will elaborate on the process by which PLAA generates specific adversarial traffic through the step-by-step generation of packet-level features. First, we introduce the selection of packet-level features in network traffic. Next, we discuss how constraints are added to ensure that the generated adversarial traffic achieves breakthroughs in both feasibility and attack semantics. Finally, we present the specific framework and traffic generation algorithm.

\subsection{Features of traffic}

As mentioned earlier, most DL-based NIDS detect malicious traffic by analyzing the features of network traffic, which are predominantly flow-level features. In this section, we will provide a detailed introduction to the attribute features and statistical features of network traffic flow and explain how packet-level features can be utilized to represent flow-level traffic characteristics.

\subsubsection{Feature selection}

The attribute features of network traffic consist of a five-tuple, namely ``Source IP'', ``Destination IP", ``Source Port", ``Destination Port", and ``Protocol". However, such features do not reflect the characteristics of the traffic itself, they primarily represent the inherent attributes of the traffic. Typically, DL-based NIDS rely on the statistical features of traffic flows as the foundation for classification. The statistical features can further be subdivided into four dimensions: Length, Count, Time and Speed. To classify the statistical features of traffic, we analyzed the feature extractors used by NIDS during the monitoring phase (using CICFlowMeter as an example). The classification of these statistical features is shown in Table \ref{table2}.

% \begin{table}[h]
% \centering
% \renewcommand{\arraystretch}{1.3} % 调整行高
% %特征分类
% \caption{The classification of these statistical features.}
% \label{table2}
% \resizebox{\columnwidth}{!}{%
% \begin{tabular}{|c|c|}
% \hline
% \textbf{Category} & \textbf{Features Name} \\ \hline
% Length & \begin{tabular}[c]{@{}c@{}}tot\_l\_pkt; pkt\_l\_max; pkt\_l\_min; \\ pkt\_l\_avg; pkt\_l\_std; hdr\_len\end{tabular} \\ \hline
% Count & \multicolumn{1}{l|}{\begin{tabular}[c]{@{}l@{}}tot\_pk; psh\_flag; urg\_flag; fin\_cnt; syn\_cnt; rst\_cnt;\\ pst\_cnt; ack\_cnt; urg\_cnt; cwe\_cnt; ecc\_cnt\end{tabular}} \\ \hline
% Time & fl\_dur; iat\_max; iat\_min; iat\_avg; iat\_std; iat\_tot \\ \hline
% Speed & byt\_s;pkt\_s \\ \hline
% \end{tabular}%
% }
% \end{table}

\begin{table}[h]
\centering
\renewcommand{\arraystretch}{1.3} % 调整行高
% 需要先在文档导言区添加 \usepackage{booktabs}
\caption{The classification of these statistical features.}
\label{table2}
\resizebox{\columnwidth}{!}{%
\begin{tabular}{l l} % 移除竖线，两列均左对齐
\toprule % 顶部粗线
\textbf{Category} & \textbf{Features Name} \\
\midrule % 中部细线（表头与数据分隔）
Length & \begin{tabular}[c]{@{}l@{}}tot\_l\_pkt; pkt\_l\_max; pkt\_l\_min; \\ pkt\_l\_avg; pkt\_l\_std; hdr\_len\end{tabular} \\
Count & \begin{tabular}[c]{@{}l@{}}tot\_pk; psh\_flag; urg\_flag; fin\_cnt; syn\_cnt; rst\_cnt;\\ pst\_cnt; ack\_cnt; urg\_cnt; cwe\_cnt; ecc\_cnt\end{tabular} \\
Time & fl\_dur; iat\_max; iat\_min; iat\_avg; iat\_std; iat\_tot \\
Speed & byt\_s; pkt\_s \\
\bottomrule % 底部粗线
\end{tabular}%
}
\end{table}

To represent the four types of flow-level statistical features through packet-level features, the packet-level feature $P_{i}$ is defined as 
\begin{align}\label{eq3}
{P}_{i}=\left\{{h}_{i}, {p}_{i}, {t}_{i}, {f}_{i}\right\}
\end{align}

The four basic features: ${h}_{i}, {p}_{i}, {t}_{i}, {f}_{i}$, respectively represented as ``Header Length", ``Payload Length", ``Inter-Packet Interval" and ``Flags" information of the packet $P_{i}$.

\subsubsection{Feature representation}

Considering that traffic features have different data types and ranges, and existing literature has highlighted the significant heterogeneity of features across different types of traffic \cite{nasr2017compressive}, directly applying raw data without proper processing may degrade the performance of adversarial attacks. Therefore, it is crucial to adopt appropriate data processing techniques to effectively represent these features.

The packet-level statistical features of traffic $P_{i}$ illustrate the specific behavior of network traffic, directly indicating whether the traffic is benign or malicious. Due to their heterogeneous nature, we apply normalization or one-hot encoding based on the feature type. For example, the Header Length $h_{i}$, which ranges from 20 to 60 bytes, is constrained by fixed formats and options, and can only grow in multiples of 4 bytes. Based on this, one-hot encoding is more suitable for representing this feature. For the Payload Length $p_{i}$, its maximum length is 65535 bytes, making normalization a more appropriate choice for encoding this feature. For the Inter-Packet Interval $t_{i}$, it is typically measured in a range from a few milliseconds to several hundred milliseconds in practical scenarios. However, theoretically, $t_{i}$ has no upper limit. In the dataset, we observed that the maximum $t_{i}$ reached up to 120 seconds. Therefore, normalization is applied to encode this feature. For the Flags $f_{i}$, one-hot encoding is used for representation. The detailed encoding schemes for these features are summarized in Table \ref{table3}.

% \begin{table}[h]
% \centering
% \renewcommand{\arraystretch}{1.3} % 调整行高
% %特征分类
% \caption{Feature Encoding Classification.}
% \label{table3}
% \resizebox{\columnwidth}{!}{%
% \begin{tabular}{|c|c|c|}
% \hline
% \textbf{Features} & \textbf{Description} & \multicolumn{1}{l|}{\textbf{Encoding Scheme}} \\ \hline
% $h_{i}$ & Header Length & one-hot \\ \hline
% $p_{i}$ & Payload Length & normalization \\ \hline
% $t_{i}$ & Inter-Packet Interval & normalization \\ \hline
% $f_{i}$ & Flags & one-hot \\ \hline
% \end{tabular}%
% }
% \end{table}

\begin{table}[h]
\centering
\renewcommand{\arraystretch}{1.3} % 调整行高
% 需要先在文档导言区添加：\usepackage{booktabs}
\caption{Feature Encoding Classification.}
\label{table3}
\resizebox{\columnwidth}{!}{%
\begin{tabular}{l l l} % 移除所有竖线，三列均左对齐
\toprule % 顶部粗线
\textbf{Features} & \textbf{Description} & \textbf{Encoding Scheme} \\
\midrule % 表头与数据分隔线
$h_{i}$ & Header Length & one-hot \\
$p_{i}$ & Payload Length & normalization \\
$t_{i}$ & Inter-Packet Interval & normalization \\
$f_{i}$ & Flags & one-hot \\
\bottomrule % 底部粗线
\end{tabular}%
}
\end{table}

\subsection{Attack semantic preservation}\label{Attack Semantic Preservation}

By directly generating packet-level features, we fundamentally avoid the existence issues caused by directly generating or modifying flow-level statistical features. At the same time, the packet-level features that need to be modified are constrained based on network protocol knowledge, ultimately ensuring that the generated traffic can exist realistically. However, another critical concern related to the validity of the generated traffic is whether the attack semantics of the original malicious traffic can be preserved after undergoing adversarial processing.

Existing studies suggest that different types of network attacks have distinct attack semantics, which necessitate a classification-based design of methods to preserve these semantics based on the type of network attack \cite{wang2024progen}. To address this, we conducted a comprehensive analysis of the attack types present in the CIC-UNSW-NB15, CIC-DDoS2019, and CIC-IDS-2017 datasets and defined the requirements for preserving their attack semantics. The detailed requirements are shown in Table \ref{table4}.

\begin{table}[t]
\centering
\renewcommand{\arraystretch}{1.3} 
\caption{Requirements to preserve attack semantics.}
\label{table4}
\resizebox{\columnwidth}{!}{%
\begin{tabular}{l l l l} % 移除所有竖线，四列均左对齐
\toprule % 顶部粗线
\textbf{Dataset} & \textbf{Attack name} & \textbf{Description} & \textbf{Requirements} \\
\midrule % 表头与数据分隔线
 & {\color[HTML]{2D2D2D} \begin{tabular}[c]{@{}l@{}}DoS slowloris\\ \&Slowhttptest\end{tabular}} & slow DoS attack & \begin{tabular}[c]{@{}l@{}}low rate;\\ flags;\\ fixed content\end{tabular} \\
 & {\color[HTML]{2D2D2D} DoS Hulk} & fast DoS attack & \begin{tabular}[c]{@{}l@{}}high rate;\\ flags;\\ fixed content\end{tabular} \\
 & {\color[HTML]{2D2D2D} Heartbleed} & malicious requests & fixed content \\
 & {\color[HTML]{2D2D2D} Brute Force} & password cracking & \begin{tabular}[c]{@{}l@{}}high rate;\\ fixed content\end{tabular} \\
 & {\color[HTML]{2D2D2D} XSS} & script injection & fixed content \\
\multirow{-8}{*}{CIC17} & {\color[HTML]{2D2D2D} Sql Injection} & malicious requests & fixed content \\
\midrule % 添加组间分隔线
 & {\color[HTML]{333333} Fuzzers} & randomly requests & \begin{tabular}[c]{@{}l@{}}high rate;\\ fixed content\end{tabular} \\
 & {\color[HTML]{333333} Exploits} & malicious requests & fixed content \\
\multirow{-3}{*}{UNSW15} & {\color[HTML]{333333} Generic} & password cracking & \begin{tabular}[c]{@{}l@{}}high rate;\\ fixed content\end{tabular} \\
\midrule % 添加组间分隔线
 & Probe & Port/Network Scan & \begin{tabular}[c]{@{}l@{}}high rate;\\ fixed content\end{tabular} \\
 & U2R & malicious requests & fixed content \\
\multirow{-3}{*}{NSLKDD} & R2L & \begin{tabular}[c]{@{}l@{}}Port/Network Scan;\\ password cracking;\\ script injection\end{tabular} & \begin{tabular}[c]{@{}l@{}}high rate;\\ fixed content\end{tabular} \\
\bottomrule % 底部粗线
\end{tabular}%
}
\end{table}

In the table, we analyze the requirements for preserving the attack semantics of traffic based on its statistical features. For instance, in the case of slow DoS attacks such as Slowhttppost, the attack is executed by sending incomplete HTTP messages. To maintain the attack semantics, it is necessary to retain a small amount of content within the packets, such as essential request methods like ``GET", ``POST", and ``CONNECT". Therefore, to ensure its attack semantics, the length $p_{i}$ of the newly generated or modified adversarial packet must be longer than the length of the original content, and the flags must reflect the incompleteness of the packet. Additionally, after preserving the necessary content and flags, it is also important to ensure that the traffic rate remains within a low range. Although some attacks, such as cross-site scripting (XSS), do not have strict requirements for rate and flags, ensuring that the rate stays within an appropriate range and carrying reasonable flags can improve the likelihood of avoiding detection by NIDS.

\subsection{Problem formulation}
Now, consider that the attacker has prepared a set of malicious traffic flows \( \mathcal{F}_{m} \in \iota \), where \( \mathcal{F}_{m} \) represents the initial malicious traffic. This traffic can be detected by the DL-based NIDS detection model \( D(\cdot) \), such that \( D(\mathcal{F}_{m}) = C_{m} \), where \( C_{m} \) denotes the attack type of the malicious traffic. The attacker's objective is to construct an adversarial operator \( A(\cdot) \) to evade NIDS detection, such that \( D(A(\mathcal{F}_{m})) = D(\mathcal{F}_{adv}) = C_{b} \), where \( C_{b} \) represents the classification of benign traffic.

The objective of this work is to train an appropriate packet-level adversarial operator \( A(\cdot) \). Thus, the problem can be formulated as
\begin{align}\label{eq4}
    \arg \min _{A(\cdot)} \mathcal{L}(C_{b},D(A(\mathcal{F}_{m}))
\end{align}

here, \( \mathcal{L}(\cdot) \) represents the cross-entropy loss function.

\subsection{State, action and reward function}

To solve the adversarial operator \( A(\cdot) \) and ensure that the traffic composed of packet-level features can bypass NIDS detection while satisfying the constraints on packet rate, payload length, and flags, we employ a reinforcement learning approach. By iteratively generating and modifying the packet-level features of existing traffic, we constrain the payload length within the action space and use a reward function to enforce constraints on packet rate and flags. This ultimately enables the evasion of NIDS detection.

\textbf{State}: Since NIDS uses the statistical features of traffic as input, we define the statistical features of traffic as the state, which can be represented as
\begin{align}\label{eq5}
    S_{i} &=\{ s_{i,1}, s_{i,2},\cdots s_{i,m}\}= F(P_{0}, P_{1},...,P_{i})
\end{align}
Here, $F(\cdot)$ represents the monitoring phase of the NIDS, where a total of $i$ packets are represented as flow-level statistical features, and $m$ denotes the dimensionality of the flow-level features. In this work, \( S_{i} \) will be used as a substitute for \( \mathcal{F} \) as the input to \( D(\cdot) \) in subsequent steps.

\textbf{Action}: The action refers to the basic parameters required to generate a new packet, with the purpose of modifying the packet-level features $P_{i}$ of the original traffic. Therefore, the action space also includes four parameters: ``Header Length", ``Payload Length", ``Inter-Packet Interval" and ``Flags". The action space is defined as
\begin{align}\label{eq6}
    A_{i} = \{ a_{i,h}, a_{i,p}, a_{i,t},  a_{i,f}\}
\end{align}
Since we reconstruct the packets one by one, \( A_{i} \) represents both the action at step \( i \) and the operation applied to the \( i \)-th packet.

It is worth noting that for these four different types of features, to maximize the evasion performance of the generated adversarial traffic, we reshape the ``Inter-Packet Interval" of the original packet using $a_{i,t}$ as the new packet feature. To ensure that the adversarial traffic retains the attack semantics of the original malicious traffic, we first generate new flag one-hot encodings based on the original packet flags using the action $a_{i,f}$, and take the union of the new and original flag encodings. Then, to ensure that the attribute features of the traffic and the payload of the packet remain unchanged, we generate actions $a_{i,h}$ and $a_{i,p}$ to perform header and payload padding, thereby preserving the original attack semantics.

\begin{figure*}[t]
	\includegraphics[width=\textwidth]{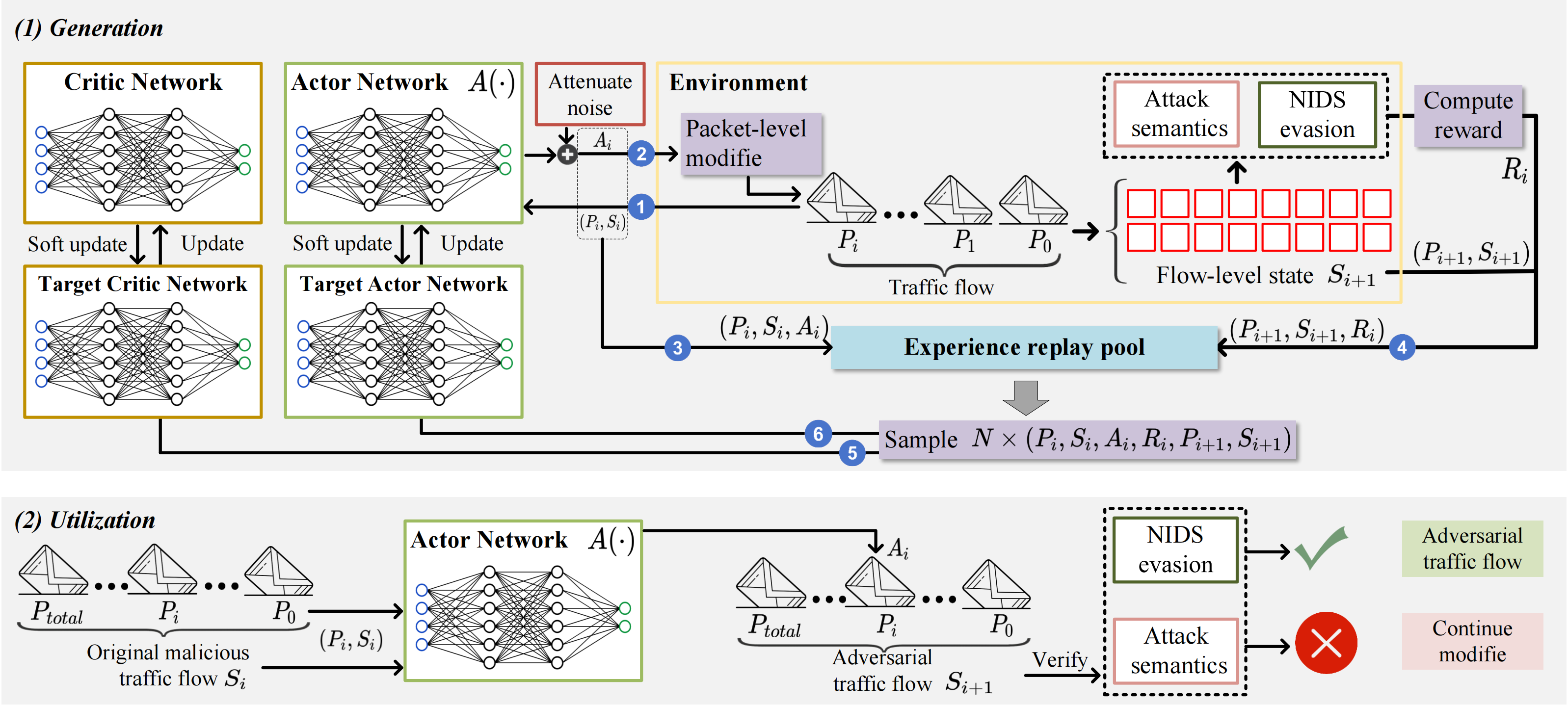}
	\caption{Adversarial attacks at the packet-level based on the actor-critic framework.} \label{fig3}
\end{figure*}

\textbf{Reward function}:
At the \( k \)-th step, after executing the action \( A_{i} \), a reward \( R_{i} \) is obtained. Typically, positive rewards are designed to encourage \( A(\cdot) \) to increase the probability of the adversarial traffic being classified as benign, while negative rewards are designed to penalize actions that cause the adversarial traffic to lose its attack semantics or lead to distortion. Thanks to the consideration of long-term rewards in RL, our objective is to generate and modify each packet within the traffic flow step by step, thereby maximizing the overall benefits of the adversarial traffic in terms of both evasion detection and attack semantics preservation. The reward function is specifically represented as follows
\begin{align}\label{eq7}
    R_{i} = \alpha _{1} r_{i}^{e}+  \alpha _{2} r_{i}^{s}
\end{align}

The reward function is divided into two parts: the first part, \( r_{i}^{e} \), incentivizes the generated adversarial traffic to evade NIDS detection, while the second part, \( r_{i}^{s} \), penalizes the loss of attack semantics in the adversarial traffic.

Specifically, for the first part, \( r_{i}^{e} = \mathcal{L}(C_{b}, D(A(\mathcal{S}_{i-1}))) - \mathcal{L}(C_{b}, D(A(\mathcal{S}_{i}))) \), where the gradual reduction in the loss function ensures that the adversarial traffic can effectively evade NIDS detection. 

The second part is expressed as:\(
r_{i}^{s} = |S_{i-1, \text{rate}} - S_{0, \text{rate}}| - |S_{i, \text{rate}} - S_{0, \text{rate}}| + \frac{\alpha_{3}}{\alpha_{2}}(\frac{1}{i-1}\sum^{i-1}_{1}|a_{i-1, h}| - \frac{1}{i}\sum^{i}_{1}|a_{i, h}|) + \frac{\alpha_{4}}{\alpha_{2}}(\frac{1}{i-1}\sum^{i-1}_{1}|a_{i-1, p}| - \frac{1}{i}\sum^{i}_{1}|a_{i, p}|)
\).
Here, \( S_{i, \text{rate}} \) and \( S_{0, \text{rate}} \) represent the traffic rate states of the current adversarial traffic and the original malicious traffic, \( \alpha \) represents the proportional coefficient.
The term \( |S_{i-1, \text{rate}} - S_{0, \text{rate}}| - |S_{i, \text{rate}} - S_{0, \text{rate}}| \) measures the difference in traffic rate between the adversarial traffic and the original malicious traffic. Maintaining a similar traffic rate to the original malicious traffic is crucial for preserving attack semantics. 
The terms \( \frac{1}{i-1}\sum^{i-1}_{1}|a_{i-1, h}| - \frac{1}{i}\sum^{i}_{1}|a_{i, h}| \) and \( \frac{1}{i-1}\sum^{i-1}_{1}|a_{i-1, p}| - \frac{1}{i}\sum^{i}_{1}|a_{i, p}| \) measure the similarity between the adversarial traffic and the original traffic in terms of header and payload characteristics. This contributes to ensuring that the adversarial traffic remains more similar to real-world traffic, enhancing its plausibility and semantic preservation.

\subsection{Generation and utilization of the adversarial operator}

The objective of the adversarial operator \( A(\cdot) \) is to output the corresponding action \( A_{i} \) at each step \( i \) based on the current feature state of the traffic, \( S_{i} \), thereby modifying each packet \( P_{i} \). Once the feature state of the traffic successfully evades NIDS detection while preserving the attack semantics, the generation process stops. The algorithm workflow is illustrated in Fig. \ref{fig3}.

During the generation phase of the adversarial operator \( A(\cdot) \), we use an Actor-Critic framework to achieve the stated objective. The RL agent utilizes two sets of neural networks: the actor network \( \mu \) and the critic network \( Q \), along with their corresponding target networks, the target actor network \( \mu' \) and the target critic network \( Q' \). The actor network serves as the adversarial operator \( A(\cdot) \) that we aim to generate. The RL agent, based on the current flow-level feature state \( S_{i} \) and the packet-level features that need to be modified, generates the action \( A_{i} \), which is shown in steps 1 and 2 in Fig. \ref{fig3}.

\begin{eqnarray}\label{eq8}
	A_{i}=A(P_{i}, S_{i}) = \mu(P_{i}, S_{i})
\end{eqnarray}

After taking an action to generate a packet, the state of the malicious traffic is updated to \( S_{i+1} \). Based on the classification probabilities provided by the NIDS and the corresponding flow-level features, the reward \( R_{i} \) is calculated. The tuple \((P_{i}, S_{i}, A_{i}, R_{i},  P_{i+1}, S_{i+1}) \) is recorded in the replay buffer, which is shown in steps 3 and 4 in Fig. \ref{fig3}. Using the experience replay method, samples are randomly drawn from the replay buffer and trained using the strategy gradient$\bigtriangledown_{\theta}J$ calculated by the critic network. It is worth noting that \( P_{i+1} \) represents the unprocessed features of the next packet-level instance, whereas \( P_{i} \), after being processed by the action \( A_{i} \), is transformed into \( P'_{i} \).
\begin{eqnarray}\label{eq9}
	\bigtriangledown_{\theta}J=\frac{1}{N} \sum_{i=1}^{N} \bigtriangledown_{\theta}\mu_{\theta}(P_{i},S_{i})\bigtriangledown_{A}Q_{\omega}(P_{i},S_{i},A_{i})
\end{eqnarray}

Here, \( N \) represents the number of samples extracted from the replay buffer, and \( \theta \) and \( \omega \) denote the parameters of the actor network and critic network, respectively. \( \gamma \) measures the importance of future rewards; the closer it is to 1, the more emphasis is placed on future rewards. 
Using the target critic network, \( y \) is computed as:

\begin{eqnarray}\label{eq10}
y = R_{i} + \gamma Q'_{\omega}(S_{i+1}, \mu'_{\theta}(P_{i+1}, S_{i+1}))
\end{eqnarray}

Then, the critic network evaluates the Q-value, and its parameters are updated by minimizing the loss function \( L \):

\begin{eqnarray}\label{eq11}
L = \frac{1}{N} \sum_{i=1}^{N} \left( y-Q_{\omega}(P_{i+1}, S_{i}, A_{i})  \right)^2
\end{eqnarray}

Based on the above formulas, the corresponding algorithm workflow can be represented as Algorithm \ref{alg1}, on lines 10 and 11 of the Algorithm \ref{alg1}, as shown in steps 5 and 6 in Fig. \ref{fig3}. The random noise \( \mathcal{N} \) introduces exploration possibilities for the agent, and \( \tau \) represents the learning rate.

The Actor Network \( \mu \) trained during the Generation phase serves as the adversarial operator \( A(\cdot) \). In the Utilization phase, actions \( A_{i} \) are generated based on Equation (\ref{eq8}), and the flow-level features of the entire traffic are updated accordingly. The traffic is then checked to determine whether it can effectively evade NIDS detection while retaining its attack semantics. If it fails to meet these criteria, the process is repeated until the desired outcome is achieved.
% \addtolength{\topmargin}{0.09 in}
% \vspace{1.5\baselineskip}
\begin{algorithm} [b!]
	\caption{Adversarial Generation Training} 
	\label{alg1}
	\begin{algorithmic}[1]
		%\REQUIRE $S_{0}, \theta, \mu,\pi(), \gamma,\ \alpha, \beta$ 
		%\ENSURE $ \theta', \mu'$ 
		\STATE Initialize the actor network $\mu_{\theta}$, critic network $Q_{\omega}$, target actor network $\mu'_{\theta}$, target critic network $Q'_{\omega}$, experience replay pool $R$
		\FOR{each episode}
		\STATE Initialize the initial traffic status $S_{0}$
		\STATE Action noise $\mathcal{N}$ attenuation
		\FOR{step $i=1 \to T_{max}$ }
		\STATE Choose action $A_{i}=\mu(P_{i},S_{i})+\mathcal{N}$
		\STATE Execute action $A_{i}$, obtain reward $R_{i}$ according to Equation (\ref{eq7}), update the environment status to $S_{i}$
		\STATE Record $(P_{i}, S_{i}, A_{i}, R_{i},  P_{i+1}, S_{i+1}))$ in replay buffer
		\STATE Sample $N$ tuples from replay buffer \\ 
        $\{ (P_{i}, S_{i}, A_{i}, R_{i},  P_{i+1}, S_{i+1}) \}_{i=1,\cdots, N}$
		\STATE Train critic network $Q$ by minimizing   Eq. (\ref{eq11})
		\STATE Train actor network $\mu$ by Eq. (\ref{eq9})
		\STATE Update the target networks \\
        $\omega' \gets \tau\omega+(1-\tau)\omega'$ \\
        $\theta' \gets \tau\theta+(1-\tau)\theta'$
		\ENDFOR
		\ENDFOR
	\end{algorithmic} 
\end{algorithm}

\section{Experiment}\label{sec5}

In this study, our goal is to design an adversarial attack method targeting DL-based NIDS. The main innovation of this method lies in starting from packet-level features to ultimately modify flow-level features. This ensures that the adversarial traffic can exist realistically and retain its attack semantics while effectively evading NIDS detection. To evaluate the method, we assess its performance from two aspects: the evasion effectiveness of the adversarial traffic and its realism and preservation of attack semantics.

 % In this section, we first introduce the basic experimental setup. Then, based on existing common NIDS, we validate that PLAA has strong capabilities in evading NIDS detection. Finally, we verify the realism and attack semantics preservation of the generated adversarial traffic by defining an Anomaly score metric.

\subsection{Experimental Setup}

% In the experimental phase, we employed three DL-based NIDS models and four ML-based NIDS models to evaluate our proposed adversarial method on three datasets. In this section, we provide a detailed overview of the datasets used and present the performance of the selected NIDS models on these datasets.
\addtolength{\topmargin}{0.01in}
\subsubsection{Datasets}

We utilized the CIC-IDS2017 \cite{sharafaldin2018toward}, CIC-UNSW-NB15 \cite{10788064}, and NSL-KDD 
 \cite{tavallaee2009detailed} datasets, including their original PCAP data and statistical data. Due to the insufficient sample sizes of certain attack types and the fact that attacks like fast DoS do not require evasion of NIDS detection, we selected specific attack types for each dataset: slow DoS attacks and brute force attacks from the CIC-IDS2017 dataset, fuzzers and generic attacks from the CIC-UNSW-NB15 dataset, and probe attacks from the NSL-KDD dataset. Additionally, since the flow-level features extracted vary across datasets, different NIDS models were trained for each dataset to ensure compatibility.

\subsubsection{NIDS}

\begin{table}[h]
\centering
\setlength{\tabcolsep}{3pt} 
\renewcommand{\arraystretch}{1.3} 
\caption{The performance of NIDS on different dataset.}
\label{table5}
\resizebox{\columnwidth}{!}{%
\begin{tabular}{ccccccc}
\hline
Dataset & \multicolumn{6}{c}{CIC-IDS2017} \\ \cline{2-7} 
Metrics & \textbf{DNN} & \textbf{CNN-LSTM} & \textbf{SVM} & \textbf{LR} & \textbf{KNN} & \textbf{RF} \\ \hline
\textbf{Accuracy} & 0.9597 & 0.9768 & 0.9607 & 0.9665 & 0.9767 & 0.9839 \\
\textbf{Precision} & 0.9588 & 0.9798 & 0.9592 & 0.9680 & 0.9798 & 0.9818 \\
\textbf{Recall} & 0.9607 & 0.9736 & 0.9623 & 0.9649 & 0.9734 & 0.9861 \\
\textbf{F1 Score} & 0.9597 & 0.9767 & 0.9608 & 0.9664 & 0.9766 & 0.9839 \\ \hline
Dataset & \multicolumn{6}{c}{CIC-UNSW-NB15} \\ \cline{2-7} 
Metrics & \textbf{DNN} & \textbf{CNN-LSTM} & \textbf{SVM} & \textbf{LR} & \textbf{KNN} & \textbf{RF} \\ \hline
\textbf{Accuracy} & 0.9477 & 0.9546 & 0.9454 & 0.9501 & 0.9558 & 0.9529 \\
\textbf{Precision} & 0.9499 & 0.9408 & 0.9485 & 0.9525 & 0.9521 & 0.9503 \\
\textbf{Recall} & 0.9453 & 0.9702 & 0.9418 & 0.9473 & 0.9599 & 0.9562 \\
\textbf{F1 Score} & 0.9476 & 0.9553 & 0.9452 & 0.9499 & 0.9561 & 0.9531 \\ \hline
Dataset & \multicolumn{6}{c}{NSL-KDD} \\ \cline{2-7} 
Metrics & \textbf{DNN} & \textbf{CNN-LSTM} & \textbf{SVM} & \textbf{LR} & \textbf{KNN} & \textbf{RF} \\ \hline
\textbf{Accuracy} & 0.8808 & 0.8271 & 0.9293 & 0.9264 & 0.9381 & 0.9551 \\
\textbf{Precision} & 0.8816 & 0.8232 & 0.9256 & 0.9242 & 0.9339 & 0.9224 \\
\textbf{Recall} & 0.8796 & 0.8329 & 0.9337 & 0.9291 & 0.9429 & 0.9499 \\
\textbf{F1 Score} & 0.8806 & 0.8281 & 0.9296 & 0.9266 & 0.9384 & 0.9361 \\ \hline
\end{tabular}%
}
\end{table}

\begin{figure*}[!h]
	\includegraphics[width=\textwidth]{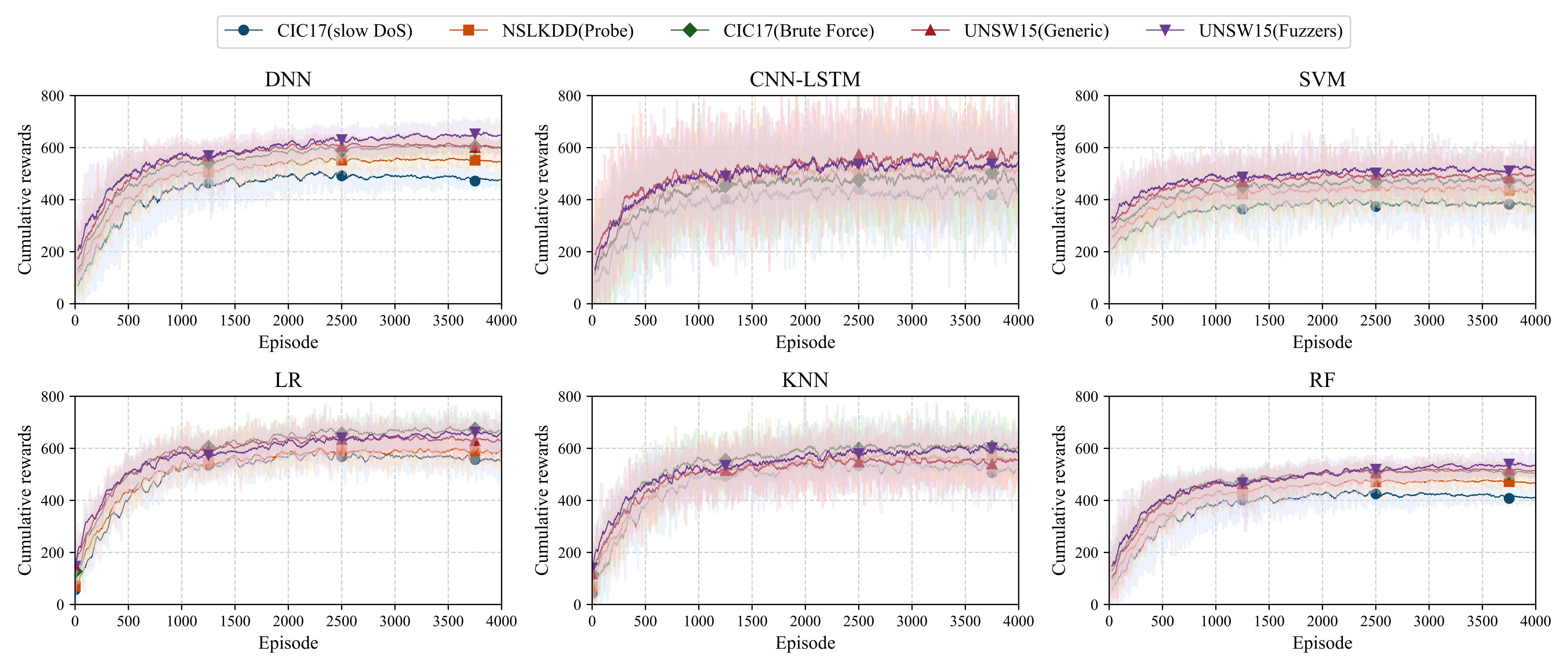}
	\caption{Comparison of convergence of different attacks under different NIDS.} \label{fig4}
\end{figure*}

The objective of PLAA is to evade NIDS detection. To evaluate its performance, we used DNN and CNN-LSTM to train two types of DL-based NIDS detection models. Additionally, we employed Support Vector Machine (SVM), Logistic Regression (LR), K-Nearest Neighbors (KNN), and Random Forest (RF) to train four types of ML-based NIDS detection models. These six NIDS detection models were trained on three datasets, and their results are presented in Table \ref{table5}.

It can be observed that on the NSL-KDD dataset, the performance of DL-based NIDS models (DNN and CNN-LSTM) is inferior to that of ML-based NIDS models. This is because the CIC-IDS2017 and CIC-UNSW-NB15 datasets have 76 traffic features, whereas the NSL-KDD dataset only contains 41 traffic features. The performance of DL models tends to degrade when the dimensionality of task features decreases, making them less effective compared to traditional ML models in such scenarios.

\subsection{The effectiveness of evasion}

In this section, we will analyze the convergence behavior of PLAA under the RL framework across different datasets. Then, we will examine its evasion effectiveness based on various NIDS models and datasets.
\subsubsection{Convergence analysis}

Cumulative rewards serve as a key indicator of the convergence of RL. To verify the convergence performance of PLAA, we trained five different models on various attack types: slow DoS and Brute Force attacks from the CIC-IDS2017 dataset, Fuzzers and Generic attacks from the CIC-UNSW-NB15 dataset, and Probe attacks from the NSL-KDD dataset.

% \begin{table*}[h]
% \centering
% \setlength{\tabcolsep}{3pt} % 调整列间距大小
% \renewcommand{\arraystretch}{1.5} % 调整行高
% \caption{Performance comparison of different attacks under different NIDS.}
% \label{table8}
% \resizebox{\textwidth}{!}{%
% \begin{tabular}{ccccccccccc}
% \hline
% Attack & \multicolumn{2}{c}{CIC17(slow DoS)} & \multicolumn{2}{c}{NSLKDD(Probe)} & \multicolumn{2}{c}{CIC17(Brute Force)} & \multicolumn{2}{c}{UNSW15(Generic)} & \multicolumn{2}{c}{UNSW15(Fuzzers)} \\ \cline{2-11} 
% NIDS & \textbf{O} & \textbf{A} & \textbf{O} & \textbf{A} & \textbf{O} & \textbf{A} & \textbf{O} & \textbf{A} & \textbf{O} & \textbf{A} \\ \hline
% \textbf{DNN} & 0.051 & 0.865 & 0.118 & 0.817 & 0.058 & 0.883 & 0.064 & 0.873 & 0.070 & 0.892 \\
% \textbf{CNN-LSTM} & 0.023 & 0.921 & 0.164 & 0.806 & 0.031 & 0.926 & 0.039 & 0.930 & 0.034 & 0.925 \\
% \textbf{SVM} & 0.042 & 0.888 & 0.084 & 0.856 & 0.039 & 0.901 & 0.061 & 0.841 & 0.059 & 0.891 \\
% \textbf{LR} & 0.035 & 0.887 & 0.069 & 0.861 & 0.036 & 0.934 & 0.049 & 0.920 & 0.022 & 0.916 \\
% \textbf{KNN} & 0.024 & 0.853 & 0.077 & 0.834 & 0.028 & 0.873 & 0.047 & 0.883 & 0.054 & 0.863 \\
% \textbf{RF} & 0.016 & 0.873 & 0.047 & 0.835 & 0.022 & 0.891 & 0.051 & 0.861 & 0.039 & 0.889 \\ \hline
% \end{tabular}%
% }
% \end{table*}

Fig. \ref{fig4} shows the convergence performance of PLAA in simulating different attacks and training different NIDS. It can be observed that the cumulative reward increases slowly after 1500 episodes and approaches convergence after 4000 episodes. In the subsequent experiments, we will use the converged model with a training episode of 4000. 

However, the cumulative return of slow DoS attacks is significantly different from the other four types of attacks, and the convergence speed is also slower. This is because, based on the basic semantic properties of the initial attack, we only modify the packet header and payload by increasing their length. This modification increases the traffic rate, which contradicts the low rate nature of the attack. Although adjusting the interval between groups can reduce traffic rate, it can also severely limit the cumulative rewards that such attacks can obtain.

Furthermore, it can be observed that the cumulative reward exhibits greater fluctuation when targeting the CNN-LSTM based NIDS compared to other NIDS. This is because CNN-LSTM has more model parameters, resulting in a more complex classification boundary. As PLAA attempts to bypass this boundary, it induces greater fluctuations in the cumulative reward.

\begin{table*}[]
\centering
\setlength{\tabcolsep}{3pt} % 调整列间距大小
\renewcommand{\arraystretch}{1.5} % 调整行高
\caption{Performance comparison of different attacks under different NIDS}
\label{table8}
\resizebox{\textwidth}{!}{%
\begin{tabular}{cclclclclcl}
\hline
Attack & \multicolumn{2}{l}{CIC17(slow DoS)} & \multicolumn{2}{l}{NSLKDD(Probe)} & \multicolumn{2}{l}{CIC17(Brute Force)} & \multicolumn{2}{l}{UNSW15(Generic)} & \multicolumn{2}{l}{UNSW15(Fuzzers)} \\ \cline{2-11} 
NIDS & \textbf{O/A} & \multicolumn{1}{c}{\textbf{ER}} & \textbf{O/A} & \multicolumn{1}{c}{\textbf{ER}} & \textbf{O/A} & \multicolumn{1}{c}{\textbf{ER}} & \textbf{O/A} & \multicolumn{1}{c}{\textbf{ER}} & \textbf{O/A} & \multicolumn{1}{c}{\textbf{ER}} \\ \hline
\textbf{DNN} & 0.051/0.865 & 0.912 & 0.118/0.817 & 0.926 & 0.058/0.883 & 0.937 & 0.064/0.873 & 0.933 & 0.070/0.892 & 0.959 \\
\textbf{CNN-LSTM} & 0.023/0.921 & 0.943 & 0.164/0.806 & 0.964 & 0.031/0.926 & 0.956 & 0.039/0.930 & 0.968 & 0.034/0.925 & 0.957 \\
\textbf{SVM} & 0.042/0.888 & 0.927 & 0.084/0.856 & 0.934 & 0.039/0.901 & 0.938 & 0.061/0.841 & 0.896 & 0.059/0.891 & 0.947 \\
\textbf{LR} & 0.035/0.887 & 0.919 & 0.069/0.861 & 0.925 & 0.036/0.934 & 0.969 & 0.049/0.920 & 0.967 & 0.022/0.916 & 0.937 \\
\textbf{KNN} & 0.024/0.853 & 0.874 & 0.077/0.834 & 0.903 & 0.028/0.873 & 0.898 & 0.047/0.883 & 0.926 & 0.054/0.863 & 0.912 \\
\textbf{RF} & 0.016/0.873 & 0.887 & 0.047/0.835 & 0.876 & 0.022/0.891 & 0.911 & 0.051/0.861 & 0.907 & 0.039/0.889 & 0.925 \\ \hline
\end{tabular}%
}
\end{table*}

\subsubsection{Avoidance analysis}

To evaluate the evasion effectiveness of PLAA against different NIDS, we use only malicious traffic as test samples, resulting in a positive sample count of zero, making metrics such as Accuracy and Precision unsuitable. Therefore, we define the following metrics: \( \textbf{O} \) (Original Malicious Traffic Pass Ratio), which represents the proportion of original malicious traffic classified as benign by the NIDS within the original samples; \( \textbf{A} \) (Adversarial Malicious Traffic Pass Ratio), which represents the proportion of adversarial traffic classified as benign by the NIDS within the original samples; and \( \textbf{ER} = \textbf{A}/(1 - \textbf{O}) \) (Evasion Rate), which quantifies the effectiveness of PLAA by measuring the proportion of originally detected malicious traffic that is successfully transformed into adversarial traffic capable of evading NIDS detection. The detailed results are shown in Table \ref{table8}.

% \begin{figure*}[!h]
% 	\includegraphics[width=\textwidth]{pic3.png}
% 	\caption{Comparison of anomaly score for different attacks under different NIDS.} \label{fig6}
% \end{figure*}

% In Fig. \ref{fig5} presents the evasion performance of five simulated attack types against six NIDS models. 

\begin{table*}[]
\centering
\setlength{\tabcolsep}{3pt} % 调整列间距大小
\renewcommand{\arraystretch}{1.3} % 调整行高
\caption{Comparison of anomaly score for different attacks under different NIDS}
\label{table9}
\resizebox{\textwidth}{!}{%
\begin{tabular}{cccccc}
\hline
Attack & CIC17(slow DoS) & NSLKDD(Probe) & CIC17(Brute Force) & UNSW15(Generic) & UNSW15(Fuzzers) \\ \cline{2-6} 
NIDS & \textbf{$\overline{\mathcal{A}_{l}}$/$\overline{\mathcal{A}_{t}}$/$\overline{\mathcal{A}_{s}}$} & \textbf{$\overline{\mathcal{A}_{l}}$/$\overline{\mathcal{A}_{t}}$/$\overline{\mathcal{A}_{s}}$} & \textbf{$\overline{\mathcal{A}_{l}}$/$\overline{\mathcal{A}_{t}}$/$\overline{\mathcal{A}_{s}}$} & \textbf{$\overline{\mathcal{A}_{l}}$/$\overline{\mathcal{A}_{t}}$/$\overline{\mathcal{A}_{s}}$} & \textbf{$\overline{\mathcal{A}_{l}}$/$\overline{\mathcal{A}_{t}}$/$\overline{\mathcal{A}_{s}}$} \\ \hline
\textbf{DNN} & 0.232/0.056/0.386 & 0.453/0.497/0.273 & 0.426/0.257/0.473 & 0.389/0.473/0.207 & 0.137/0.217/0.047 \\
\textbf{CNN-LSTM} & 0.346/0.076/0.572 & 0.531/0.489/0.385 & 0.496/0.131/0.761 & 0.473/0.541/0.263 & 0.187/0.097/0.253 \\
\textbf{SVM} & 0.341/0.032/0.630 & 0.506/0.587/0.268 & 0.354/0.220/0.400 & 0.512/0.476/0.371 & 0.169/0.130/0.184 \\
\textbf{LR} & 0.206/0.064/0.327 & 0.448/0.462/0.297 & 0.393/0.0.215/0.470 & 0.387/0.274/0.392 & 0.208/0.146/0.236 \\
\textbf{KNN} & 0.293/0.098/0.444 & 0.657/0.728/0.339 & 0.493/0.306/0.521 & 0.451/0.397/0.361 & 0.192/0.084/0.277 \\
\textbf{RF} & 0.267/0.104/0.389 & 0.746/0.819/0.370 & 0.573/0.289/0.665 & 0.276/0.368/0.135 & 0.158/0.164/0.131 \\ \hline
\end{tabular}%
}
\end{table*}
\addtolength{\topmargin}{0.01in}
It can be observed that PLAA achieves a \textbf{ER} close to or exceeding 90\% across all NIDS models, successfully converting originally detectable malicious traffic into adversarial traffic capable of evading detection. The evasion performance of the slow DoS attack against KNN\&RF based NIDS is slightly lower than that of other attacks, as modifying the Header Length and Payload Length conflicts with the intrinsic low-rate characteristics of slow DoS attacks. This also corroborates the lower cumulative reward observed in Fig. \ref{fig4} for this attack type. Additionally, since traditional ML models construct less complex classification boundaries compared to DL models, the adversarial traffic generated by PLAA may struggle to effectively bypass these relatively rigid boundaries, leading to reduced performance.
\subsection{The realism of the adversarial traffic}
Although generating adversarial traffic from packet-level features and following the constraints outlined in Section \ref{Attack Semantic Preservation} ensures its existence in the real world, there is still a gap between the adversarial traffic and real traffic. To address this, we define an Anomaly Score to measure the difference between the generated adversarial traffic and the original malicious traffic, thus verifying the realism of the adversarial traffic.

\subsubsection{Anomaly score metrics}

We evaluate the realism of adversarial traffic based on three characteristics: traffic length, time, and rate, defining anomaly scores \( \mathcal{A}_{l}, \mathcal{A}_{t}, \mathcal{A}_{s} \), which is formulated as 
\begin{align}\label{eq12}
\mathcal{A}_{l}&=\frac{1}{2i}\sum  (p_{i,\mathrm{adv}} - p_{i,\mathrm{mal}})/p_{i,\mathrm{mal}} \ +  \\ 
&\ \ \ \ \ \ \ \ \ \ \ \ (h_{i,\mathrm{adv}} - h_{i,\mathrm{mal}})/h_{i,\mathrm{mal}} \nonumber
\\
  \mathcal{A}_{t} &= \frac{1}{i}\sum{|t_{i,adv} - t_{i,mal}|}/t_{i,mal}\\
  \mathcal{A}_{s} &= \frac{1}{i}\sum{|S_{i,rate}^{adv} - S_{i,rate}^{mal}|} / {S_{i,rate}^{mal}}
\end{align}
among them, \( \mathcal{A}_{l} \) measures the difference in average packet length(\( p_{i,\mathrm{adv}}, p_{i,\mathrm{mal}}\)) and header length(\( h_{i,\mathrm{adv}}, h_{i,\mathrm{mal}}\)) between adversarial and malicious traffic.  
\( \mathcal{A}_{t}  \) quantifies the difference in average inter-packet interval between adversarial and malicious traffic.  
\( \mathcal{A}_{s} \) evaluates whether the rate of adversarial traffic aligns with the semantic requirements of the attack.

\subsubsection{Analysis of realism of adversarial traffic}

Table \ref{table9} presents the average values of the three anomaly scores, \( \overline{\mathcal{A}_{l}}, \overline{\mathcal{A}_{t}}, \overline{\mathcal{A}_{s}} \), for all adversarial traffic that successfully evaded NIDS detection across five simulated attack types and six different NIDS models.

It can be observed that in the length anomaly score \( \overline{\mathcal{A}_{l}} \), the slow DoS attack remains constrained, as it must maintain its low-speed characteristic. Excessive length growth would trigger penalties in the reward function related to transmission rate. In contrast, the Probe attack is less sensitive to this metric, likely because its original attack behavior involves scanning multiple ports, where packet content length remains relatively fixed. Exploring variations in length may help this attack better bypass classification boundaries.  

For the time anomaly score \( \overline{\mathcal{A}_{t}} \), a notable difference emerges between Generic and Brute Force attacks, despite both targeting password systems. Brute Force attacks exhibit greater deviations from their original attack patterns in this metric. This may be because, although Brute Force attacks generally require high-speed execution, slowing down the attack can effectively bypass NIDS classification boundaries. Under conditions where length changes remain minimal, increasing the time interval can significantly reduce attack speed, aiding in evasion.  

Regarding the speed anomaly score \( \overline{\mathcal{A}_{s}} \), although the slow DoS attack remains constrained in the first two metrics, it still exhibits some deviation in \( \overline{\mathcal{A}_{s}} \). This is due to the modification approach, which only increases the Header Length and Payload Length of packets. As a result, the traffic rate increases, conflicting with the attack’s inherent low-speed characteristic.

It can be observed that the slow DoS attack maintains restraint in both the average length anomaly score \( \overline{\mathcal{A}_{l}} \) and the average time anomaly score \( \overline{\mathcal{A}_{t}} \). This is because the slow DoS attack requires a low transmission speed, and excessive growth in packet length would lead to penalties in the reward function related to transmission rate. However, in terms of the average speed anomaly score \( \overline{\mathcal{A}_{s}} \), there is still a significant difference from the original attack. This discrepancy arises because the modification process only increases the Header Length and Payload Length, leading to an unintended increase in transmission speed, which conflicts with the inherently low-speed characteristics of slow DoS attacks.

For the Probe attack, both the average length anomaly score \( \overline{\mathcal{A}_{l}} \) and the average time anomaly score \( \overline{\mathcal{A}_{t}} \) show considerable deviations from the original attack. This may be due to the nature of Probe attacks, which involve scanning different ports, resulting in relatively fixed packet content length and inter-packet intervals. Consequently, modifying these two attributes helps adversarial traffic better bypass classification boundaries.

Although Brute Force and Generic attacks both target password systems, they exhibit significant differences in performance. Brute Force attacks show a more substantial deviation in the average speed anomaly score \( \overline{\mathcal{A}_{s}} \) compared to the original attack. This is likely because, despite the need for high-speed execution in Brute Force attacks, slowing down their attack speed enhances their ability to evade NIDS detection.

The Fuzzers attack, in general, demonstrates superior performance compared to the other attack types, as it maintains relatively restrained modifications across all anomaly scores.

\section{Conclusion }\label{sec6}

In this study, we analyzed the two primary deficiencies ``existence\&semantics", caused by current adversarial attacks against ML\&DL-based NIDS. We also discussed the drawbacks of three existing approaches that attempt to address these issues. To overcome these limitations, we proposed the PLAA method, which employs a RL framework to generate packet-level features incrementally, assembling them into network flows to effectively resolve the existence issue of adversarial traffic. Furthermore, we incorporated an attack semantics term into the reward function to ensure that the generated adversarial traffic retains its intended attack semantics. In our experimental results, we tested our method based on existing NIDS defense models to validate its performance in evasion detection and attack semantic preservation. This study primarily focused on the theoretical methodology for generating adversarial traffic, with evaluation heavily reliant on datasets. In future work, we plan to shift our focus toward real-world malicious network traffic and explore more practical adversarial techniques.

\bibliographystyle{IEEEtran}
\bibliography{references}

\end{document}